\documentclass[aps,prd,twocolumn,groupedaddress,amssymb,eqsecnum,showpacs,epsfig,nofootinbib]{revtex4}
\usepackage{graphicx}
\usepackage{bm}
\usepackage{dcolumn}
\usepackage{amsmath}

\usepackage{amsmath}
\usepackage{amssymb}

\numberwithin{equation}{section}
\usepackage{epsfig}

\def \lleq {\lower0.9ex\hbox{ $\buildrel < \over \sim$} ~}
\def \ggeq {\lower0.9ex\hbox{ $\buildrel > \over \sim$} ~}

\def \om    {\Omega}

\def \omms   {\Omega_m}

\def \omm  {\Omega_{0 {\rm m}}}

\def \beq  {\begin{equation}}
\def \eeq  {\end{equation}}
\def \ber  {\begin{eqnarray}}
\def \eer  {\end{eqnarray}}

\newcommand {\ga} {\ {\raise-.5ex\hbox{$\buildrel>\over\sim$}}\ }
\newcommand {\la} {\ {\raise-.5ex\hbox{$\buildrel<\over\sim$}}\ }
\newcommand{\eqn}[1] {Eq.~(\ref{#1})}
\newcommand{\fig}[1] {Fig.~(\ref{#1})}
\renewcommand{\(}{\left(}
\renewcommand{\)}{\right)}

\renewcommand{\]}{\right]}

\begin{document}
\newcommand{\newc}{\newcommand}

\newc{\be}{\begin{equation}}
\newc{\ee}{\end{equation}}
\newc{\ba}{\begin{eqnarray}}
\newc{\ea}{\end{eqnarray}}
\newc{\bea}{\begin{eqnarray*}}
\newc{\eea}{\end{eqnarray*}}
\newc{\D}{\partial}
\newc{\ie}{{\it i.e.} }
\newc{\eg}{{\it e.g.} }
\newc{\etc}{{\it etc.} }
\newc{\etal}{{\it et al.}}
\newc{\lcdm }{$\Lambda$CDM }
\newcommand{\nn}{\nonumber}
\newc{\ra}{\rightarrow}
\newc{\lra}{\leftrightarrow}
\newc{\lsim}{\buildrel{<}\over{\sim}}
\newc{\gsim}{\buildrel{>}\over{\sim}}
\title{New Parametrization for the Scale Dependent Growth Function in General Relativity}
\author{James B. Dent$^a$, Sourish Dutta$^a$ and Leandros Perivolaropoulos$^b$}
 \affiliation{$^a$Department of Physics and Astronomy, Vanderbilt University,
Nashville, TN 37235 \\$^b$Department of Physics, University of Ioannina, Greece}
\date{\today}

\begin{abstract}
We study the scale dependent evolution of the growth function $\delta(a,k)$ of cosmological perturbations in dark energy models based on General Relativity. This scale dependence is more prominent on cosmological scales of $100h^{-1}Mpc$ or larger. We derive a new scale dependent parametrization which generalizes the well known Newtonian approximation result $f_0(a)\equiv \frac{d\ln \delta_0}{d\ln a}=\om(a)^\gamma$ ($\gamma =\frac{6}{11}$ for \lcdm) which is a good approximation on scales less than $50h^{-1}Mpc$. Our generalized parametrization is of the form $f(a)=\frac{f_0(a)}{1+\xi(a,k)}$ where $\xi(a,k)=\frac{3 H_{0}^{2} \omm}{a k^2}$. We demonstrate that this parametrization fits the exact result of a full general relativistic evaluation of the growth function up to horizon scales for both \lcdm and dynamical dark energy. In contrast, the scale independent parametrization does not provide a good fit on scales beyond $5\%$ of the horizon scale ($k\simeq 0.01 h^{-1}Mpc$).
\end{abstract}
\maketitle

\section{Introduction}
The observable growth function  $\delta(a)\equiv \frac{\delta \rho}{\rho}(a)$ of the linear matter density contrast as a function of the cosmic scale factor $a$ (or equivalently the redshift $z$) provides a useful tool to test theoretical models attempting to explain the observed accelerating expansion of the universe. In particular, the combination of the observed expansion rate of the universe $H(a)$ \cite{SN,CMB,BAO} with the growth function $\delta (a)$ \cite{Nesseris:2007pa,DiPorto:2007ym,Polarski:2007rr,lindercahn,Bertschinger:2006aw,Gong:2008fh} can provide significant insight into the properties of dark energy \cite{Nesseris:2006er,Uzan:2006mf} driving the accelerating expansion (e.g. sound speed, existence of anisotropic stress etc) or even distinguish it from modified gravity \cite{Boisseau:2000pr} which may alternatively be responsible for the accelerating expansion.

The sub-Hubble (scale independent) growth function can be obtained for any homogeneous dark energy cosmology in the context of General Relativity by solving numerically the growth equation \cite{Dodelson:2003ft}. On larger scales, there are several numerical studies solving the full coupled relativistic equations for cosmological perturbations and making very precise predictions about structure at large-scales and/or high redshift (eg \cite{numrel}). Alternatively, the growth of cosmological perturbations may be approximated by using a simple parametrization. The standard parametrization of the linear growth function $\delta (a)$ is usually made by introducing a growth index $\gamma$ defined by  \be f_{0}(a)\equiv \frac{d\ln \delta_{0}}{d\ln a}=\omms(a)^\gamma \label{f0def}\ee where $a=\frac{1}{1+z}$ is the scale factor and
\be \omms (a) \equiv \frac{H_0^2 \omm a^{-3}}{H(a)^2} \label{omadef} \ee
 is the ratio of the matter density to the critical density when universe has scale-factor $a$ where $H_0$ is Hubble constant and  $\Omega_{0m}$ is ratio of mass density to critical density.
This parametrization \cite{Wang:1998gt} provides an excellent fit to the evolution equation for $\delta(a)$ in the small scale (sub-Hubble) approximation \be {\ddot \delta} + 2 H {\dot \delta} - 4\pi G \rho_m \delta =0 \label{greqtim} \ee  where  an overdot denotes the derivative with respect to time and $\rho_m$ is the matter density. Changing variables from $t$ to $\ln a$ we obtain the evolution equation for the growth factor $f$ as
\be f' + f^2 + f(\frac{\dot H}{H^2}+2)=\frac{3}{2} \omms \label{greqlna} \ee  where $' =d/dlna$.
For dark energy models in a flat universe with a slowly varying equation of state $w(a)\equiv \frac{p(a)}{\rho(a)}=w_0$, the solution of eq. (\ref{greqlna}) is well approximated by  eq. (\ref{f0def}) with \cite{Wang:1998gt} \be \gamma=\frac{3(w_0 -1)}{6w_0-5} \label{gamval} \ee which reduces to $\gamma=\frac{6}{11}$ for the \lcdm case ($w_0=-1$). It is therefore clear that the observational determination of the growth
index $\gamma$ can be used to test \lcdm \cite{Nesseris:2007pa}. It has been shown \cite{lindercahn} that even in the context of dynamical dark energy models consistent with Type Ia supernovae (SnIa) observations the parameter $\gamma$ does not vary by more than $5\%$ from its \lcdm value. However, in the context of modified gravity models $\gamma$ can vary by as much as $30\%$ (e.g. for the DGP model\cite{dgp} $\gamma_{DGP}\simeq 0.68$ \cite{lindercahn}) while scale dependence is also usually introduced\cite{Nesseris:2006er,Uzan:2006mf,Polarski:2007rr}.

Current observational constraints on $\gamma$ are based on redshift
distortions of galaxy power spectra \cite{Hawkins:2002sg}, the rms
mass fluctuation $\sigma_8(z)$ inferred from galaxy and
$Ly-\alpha$ surveys at various redshifts
\cite{Viel:2004bf}-\cite{Viel:2005ha}, weak lensing statistics
\cite{Kaiser:1996tp}, baryon acoustic oscillations
\cite{Seo:2003pu}, X-ray luminous galaxy clusters
\cite{Mantz:2007qh}, Integrated Sachs-Wolfe (ISW) effect
\cite{Pogosian:2005ez} etc. Unfortunately, the currently available
data are limited in number and accuracy and come mainly from the
first two categories. They involve significant error bars and
non-trivial assumptions that hinder a reliable determination of
$\gamma$. Thus, the current constraints on $\gamma$ are fairly weak \cite{Nesseris:2007pa} and are expressed as \be \gamma=0.674^{+0.195}_{-0.169}
\label{gambf}\ee This however is expected to change in the next few years when more detailed weak lensing surveys are anticipated to narrow significantly the above range.

A crucial assumption made in the derivation of eq. (\ref{greqtim}) in the context of general relativity metric perturbations is the assumption that the scale of the perturbations is significantly smaller than the Hubble scale \cite{Wang:1998gt}. This assumption however does not lead to a good approximation on scales larger than about $50h^{-1}Mpc$ \cite{Dent:2008ia}. In fact, the perturbed metric of spacetime takes the form (in the Newtonian gauge):
\begin{equation}
 ds^2 = -(1+2\Phi) dt^2 + (1-2\Phi)a^2\gamma_{ij}dx^i dx^j,
\end{equation}
where $\gamma_{ij}$ is the metric of the spatial section {and we are ignoring anisotropic stresses}. The evolution of density perturbations
on all scales is dictated by combining the background equations \ba H^2 &=& \frac{8\pi G}{3}(\rho_m +\rho_{de}) \label{bcgeq1} \\ {\dot \rho} &=& -3H (\rho + p) \label{bcgeq2} \ea    (assuming a flat universe with only
pressureless dark matter  and (non-clustering) dark energy) with the perturbed  linear order Einstein equations in the Newtonian gauge \cite{mabertschinger} ($\rho_m$ and $\rho_{de}$ are the matter and dark energy densities respectively while $p=w \rho$ is the pressure). The resulting (anisotropic stress-free) equations are of the form
\ba
\label{grper1}\ddot{\Phi}&=&-4H\dot{\Phi}+8\pi G \rho_{de} w_{de}\Phi\\
\label{grper2}\dot{\delta}&=&3\dot{\Phi}+\frac{k^2}{a^2}v_{f}\\
\label{grper3}\dot{v}_{f}&=&-\Phi
\ea
with constraint equations
\ba
\label{grcons1}3H(H\Phi+\dot{\Phi})+\frac{k^2}{a^2}\Phi&=&-4\pi G\delta\rho_m\\
\label{grcons2}(H\Phi+\dot{\Phi})&=&-4\pi G\rho_m v_{f}
\ea
where $\Phi$ is the Newtonian potential, $v_f\equiv-v a$ ($v$ is the velocity potential for dark matter) and we have generalized the derivation of Ref. \cite{Dent:2008ia} to the case of a general dark energy equation of state parameter $w_{de}(a)$. Clearly, equations (\ref{grper1})-(\ref{grper3}) involve a scale $k$ dependence in contrast to the small scale approximate equation (\ref{greqtim}) which is scale independent. It has been demonstrated \cite{Dent:2008ia} that for $w_{de}=-1$ (\lcdm), the solution of equations (\ref{grper1})-(\ref{grper3}) (with initial conditions for $\Phi$ and $v_f$ derived from the constraint equations with ${\dot \Phi}=0$) deviates significantly from the solution of the approximate equation (\ref{greqtim}) on scales larger than about $50h^{-1}Mpc$ ($k<0.02 h Mpc$). This is demonstrated in Fig. 1 where we compare the growth factor $f(z)$ based on the solution of the full general relativistic linearized equations (\ref{grper1})-(\ref{grper3}) with the corresponding growth factor of the standard parametrization (\ref{f0def}) for $k=0.001hMpc^{-1}$, $k=0.004hMpc^{-1}$ and $k=0.01hMpc^{-1}$ ($w_{de}=-1$ and $\omm=0.3$).

\begin{figure}
	\epsfig{file=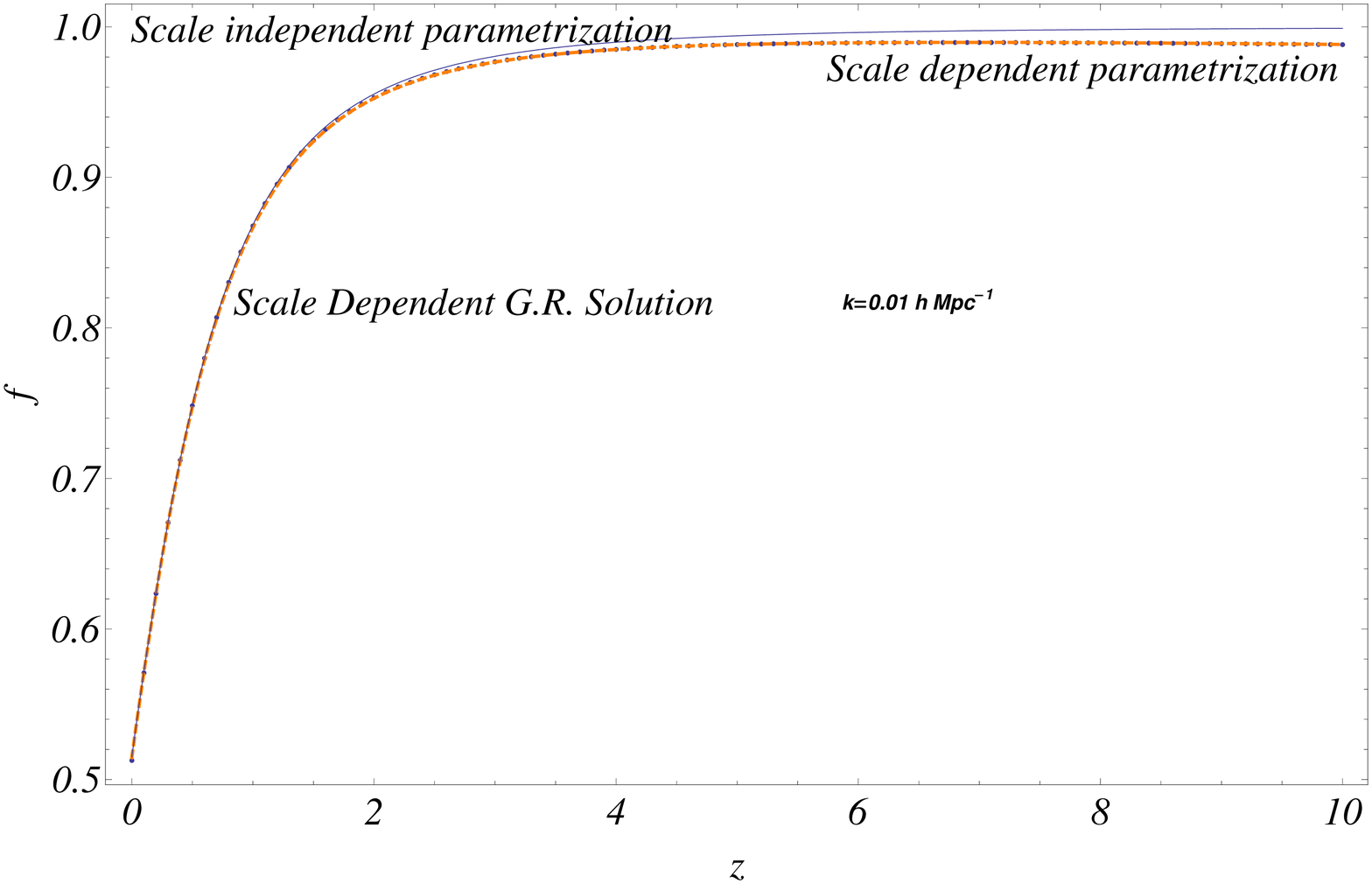,height=55mm}
	\epsfig{file=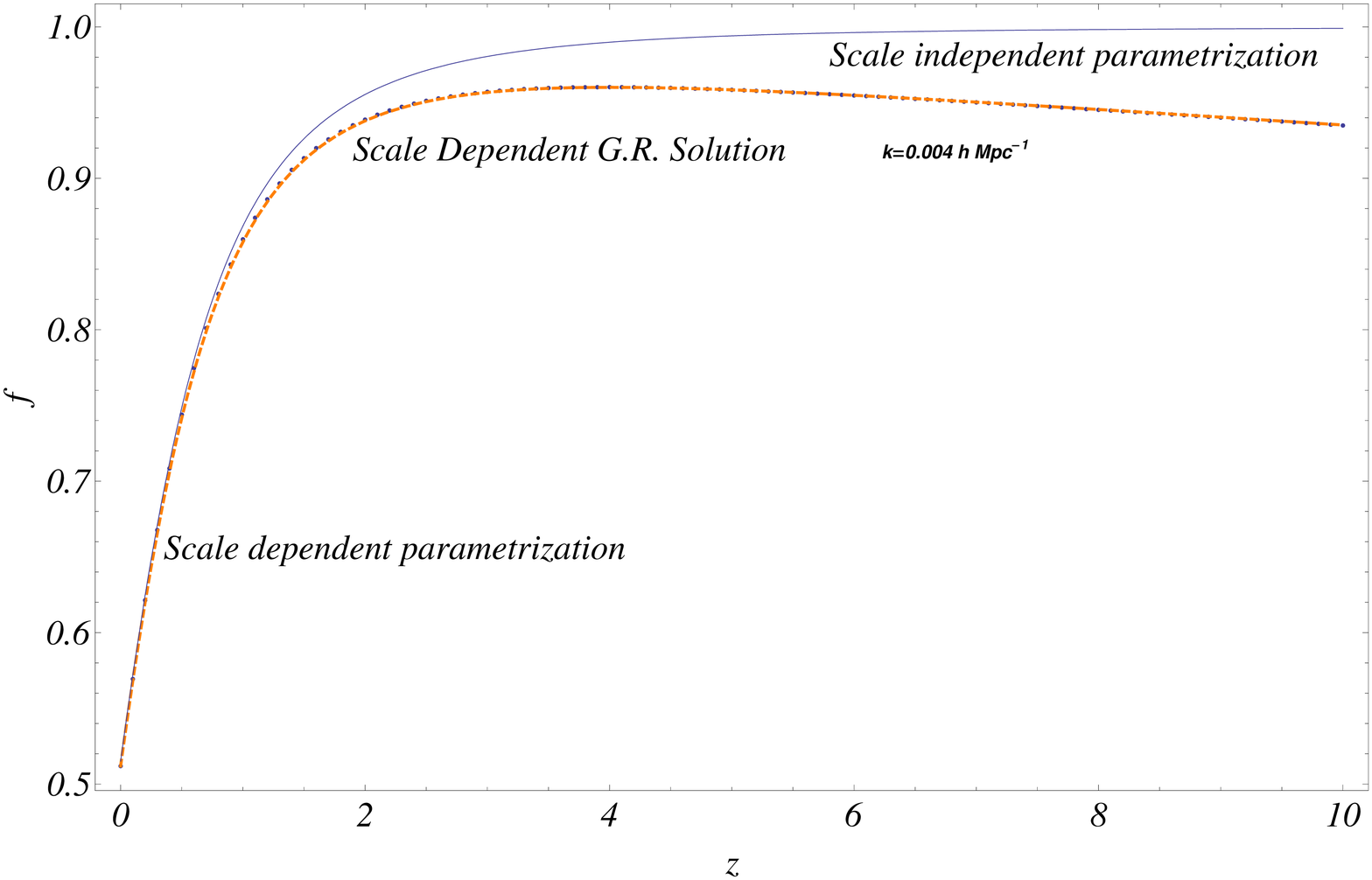,height=55mm}
	\epsfig{file=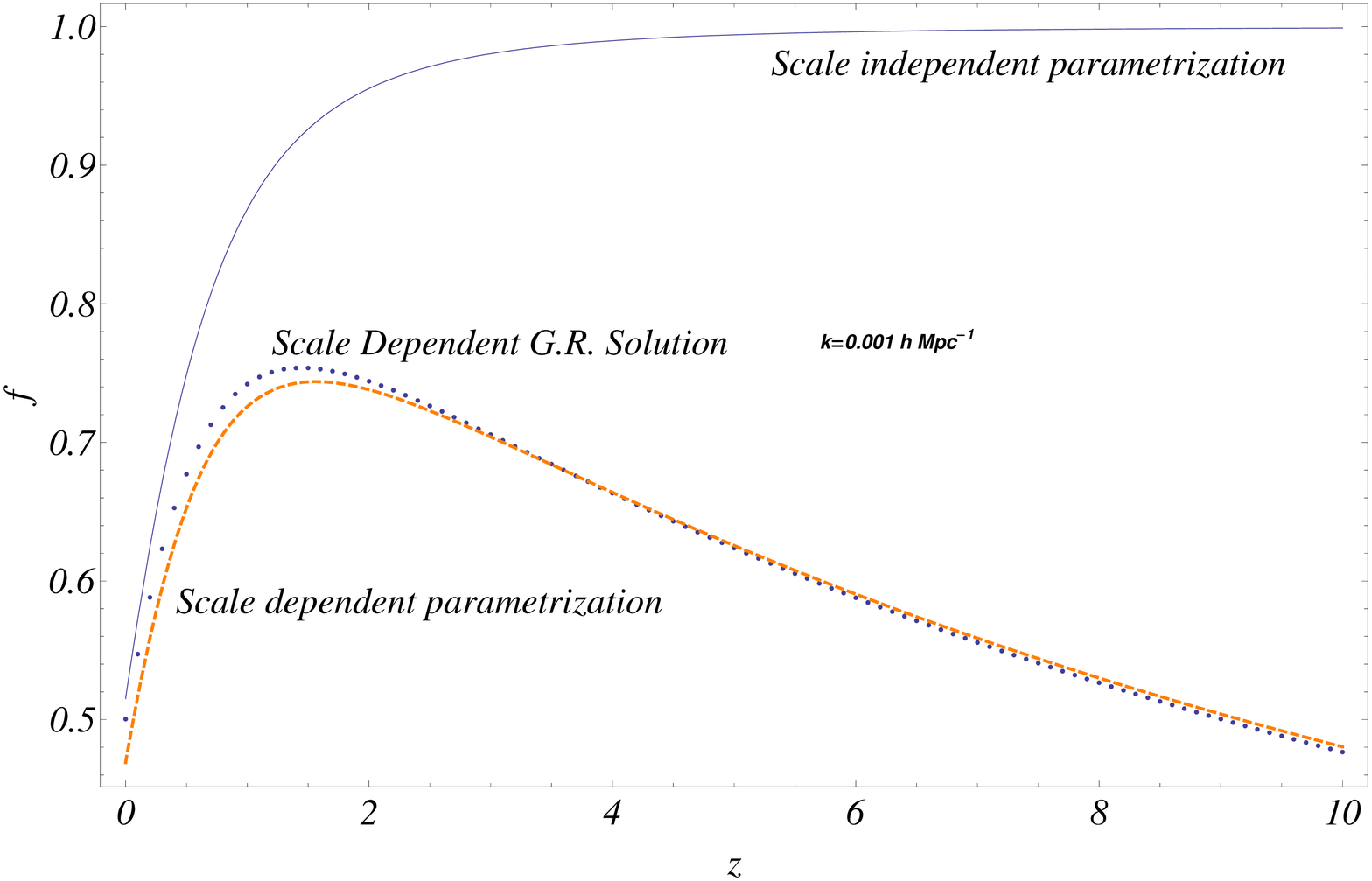,height=55mm}
	\caption
	{\label{ffits}
a: The growth factor $f$ obtained from the solution of the general relativistic system ($k=0.01h^{-1}Mpc$, $\omm=0.3$, \lcdm, dotted line) compared with the scale independent parametrization (continuous line) and the corresponding generalized scale dependent parametrization (thick dashed line) b: Similar to a. for $k=0.004h^{-1}Mpc$. c: Similar to a. for $k=0.001h^{-1}Mpc$. }
\end{figure}

\begin{figure}
	\epsfig{file=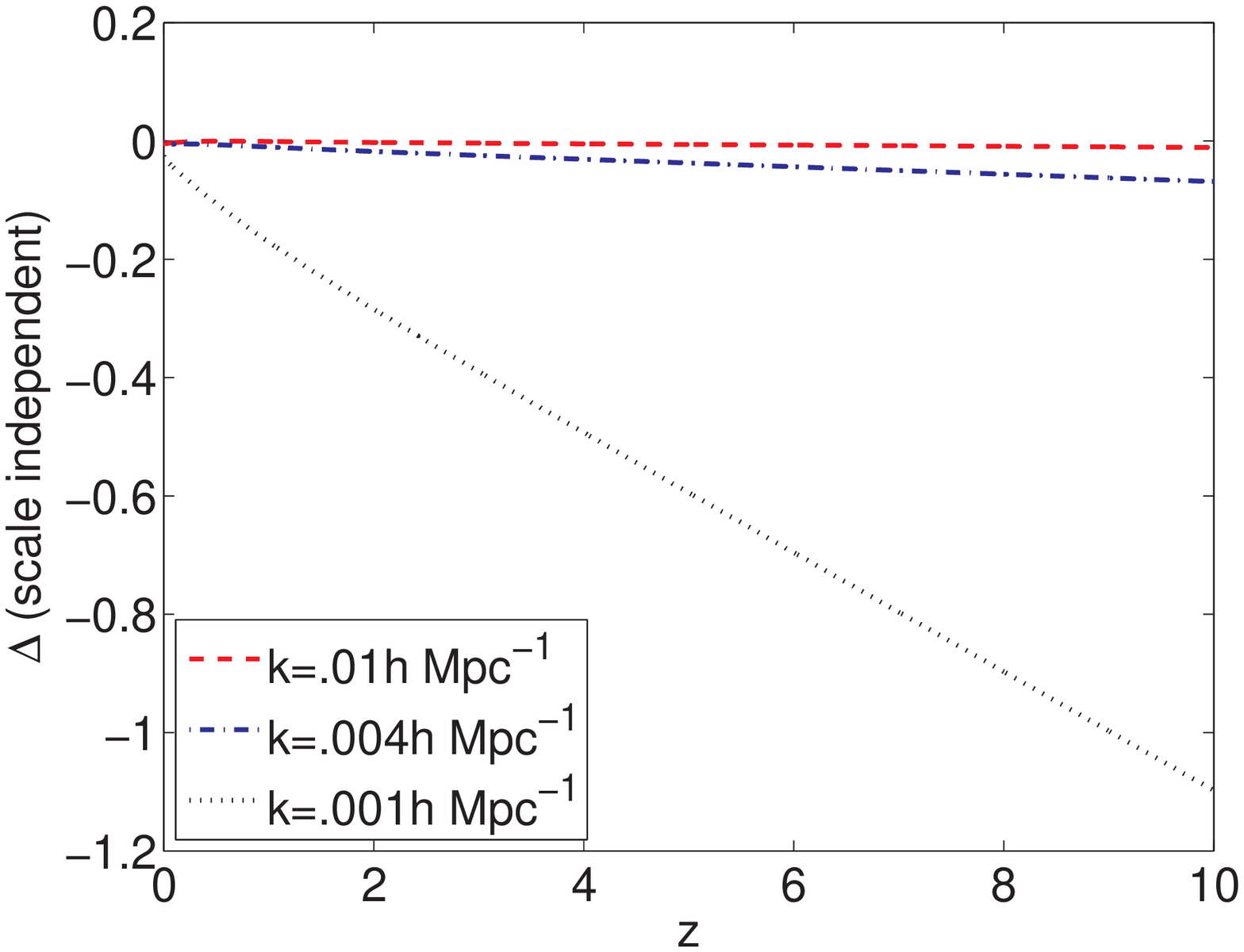,height=55mm}
	\epsfig{file=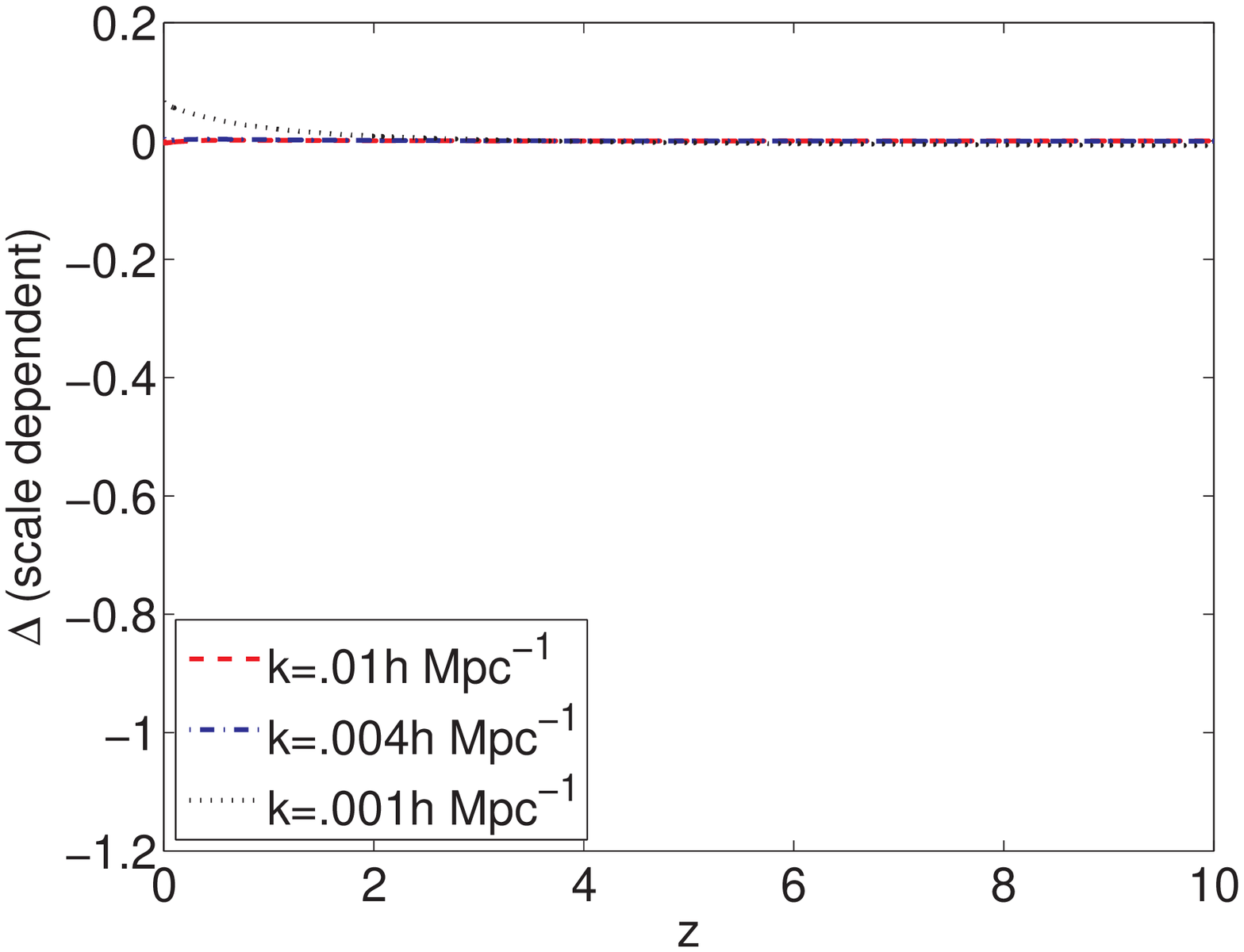,height=55mm}
	\caption
	{\label{errs}
a: The accuracy of the scale independent parametrization in terms of $\Delta$ as defined in \eqn{Delta} .  b: Likewise for the scale dependent parameterization. }
\end{figure}

\begin{figure*}[ht]
\centering
\begin{center}
$\begin{array}{@{\hspace{-0.10in}}c@{\hspace{0.0in}}c}
\multicolumn{1}{l}{\mbox{}} &
\multicolumn{1}{l}{\mbox{}} \\ [-0.2in]
\epsfxsize=3.3in
\epsffile{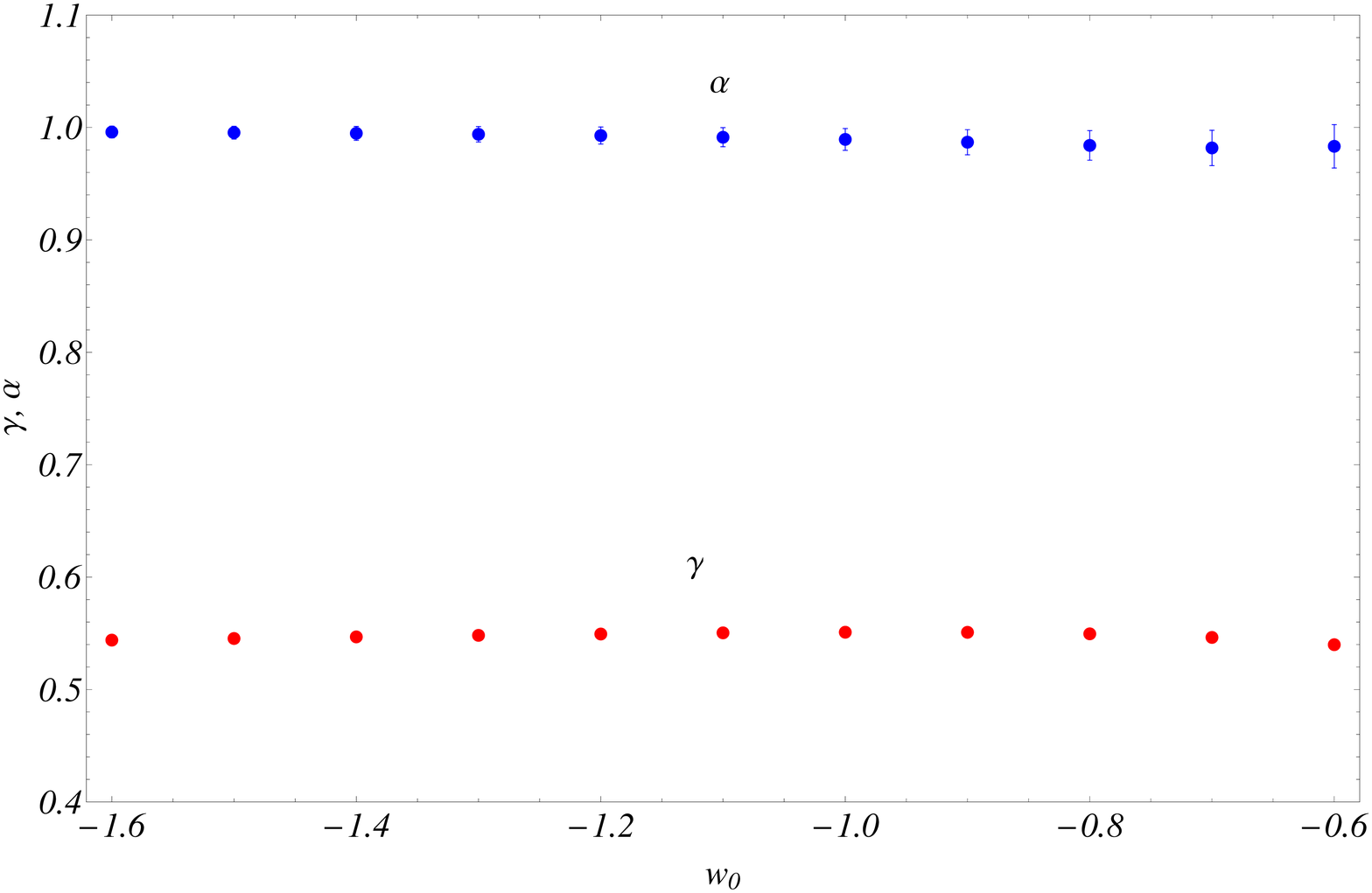} &
\epsfxsize=3.3in
\epsffile{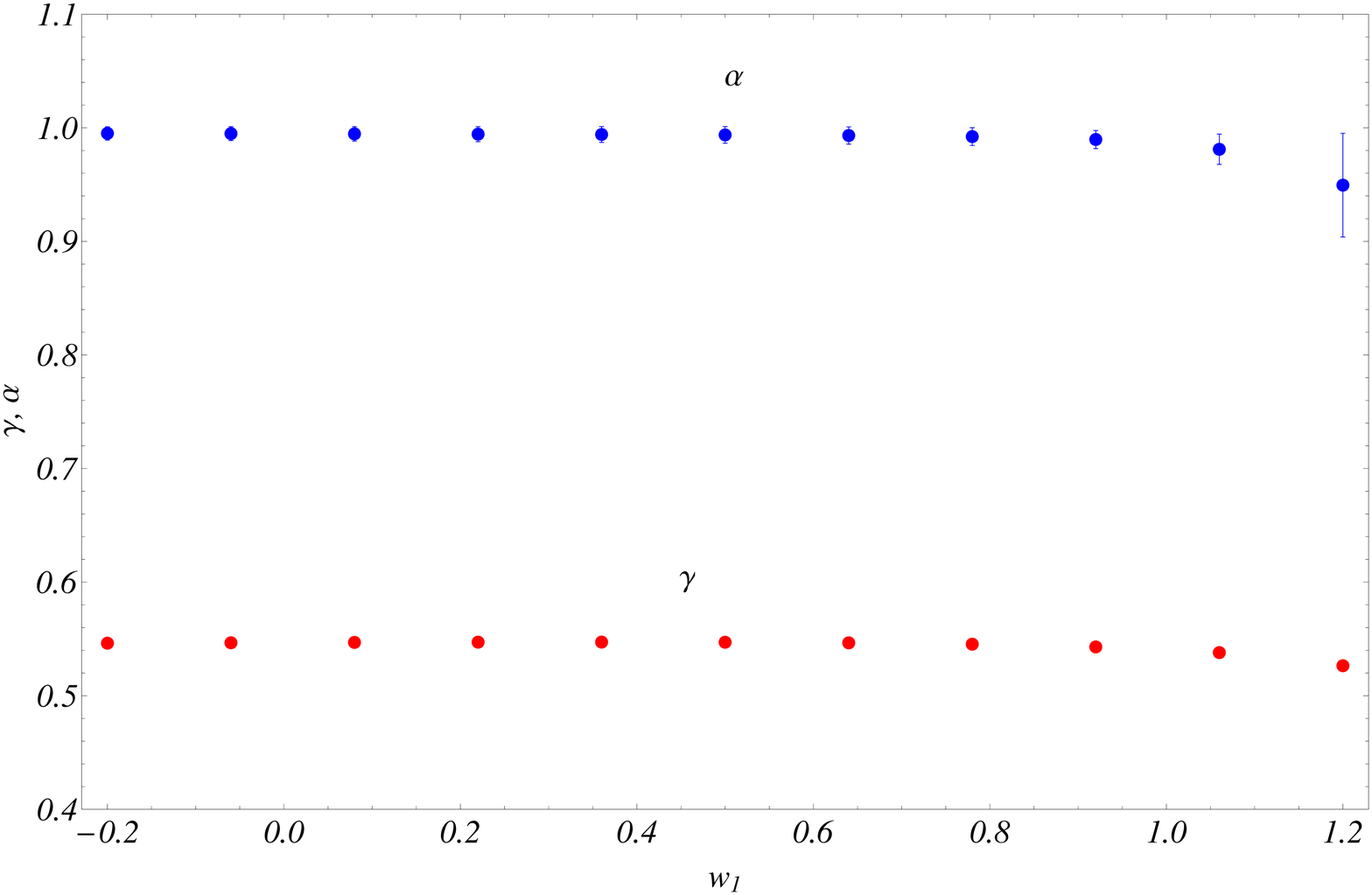} \\
\end{array}$
\end{center}
\vspace{0.0cm}
\caption{\small a: The values of the parameters $(\gamma,\alpha)$ that fit best the general relativistic growth solution as functions of the dark energy equation of state parameter $w_0$ ($w_1$ was fixed to 0). Blue disks correspond to $\alpha$, black triangles to $\gamma$. The error bars in $\alpha$ describe the $1\sigma$ variation of $\alpha$ as the scale $k$ is varied in the range $k\in [0.001,0.1]h^{-1}Mpc$ b: Similar to a for the dark energy equation of state parameter $w_1$ ($w_0$ was fixed to -1.4).}
\label{fig2}
\end{figure*}

Since this range of scales is accessible to present and especially to future large scale structure observations, it becomes clear that there is a need for an improved parametrization for the growth factor $f$ that will be compatible with the full general relativistic solution up to horizon scales. The derivation of such a parametrization is the goal of the present study\footnote{Note that if one works in the synchronous gauge, then \eqn{greqtim} is exact (see e.g. the discussion in \cite{Dent:2008ia}). However, the focus of this work is the Newtonian gauge.}.

The structure of this paper is as follows: In the next section we focus on the case of \lcdm and show that there is a scale dependent generalization of the parametrization (\ref{f0def}) that approximates well the full general relativistic linear growth solution up to horizon scales ($k\simeq 10^{-3} h Mpc^{-1}$). In section III we generalize this result to the case of dynamical dark energy models and finally in section IV we conclude and point out future extensions of this work.

\section{Scale Dependent Growth Parametrization in \lcdm}

We start by sketching the derivation of equation (\ref{greqtim}) in the context of general relativity using the  sub-Hubble approximation. The linear matter overdensity $\delta \rho_m$ may be expressed \cite{mabertschinger} in terms of the gravitational potential $\Phi$ and the background variables as follows: \be
-4\pi G\delta\rho = \frac{k^2}{a^2}\Phi +3H^2\Phi +3H\dot{\Phi} \label{drhom}
\ee
In the sub-Hubble (small scale) approximation ($\frac{k^2}{a^2} \gg H^2$) equation (\ref{drhom}) takes the form
\be
-4\pi G\delta\rho = \frac{k^2}{a^2}\Phi \label{drhomss}
\ee where we have also assumed a slowly varying gravitational potential $\Phi$.  On dimensional grounds we would expect that $3 H^2 \Phi \simeq 3 H {\dot \Phi}$ and thus it may seem unnatural to drop the $3 H {\dot \Phi}$ while keeping the $3 H^2 \Phi$. For example the effective combination of both terms would be $6 H^2 \Phi$. However, as discussed in section 3 where an arbitrary coefficient is considered for the term $H^2 \Phi$, the general relativistic solution is best approximated by a term of the form $3 H^2 \Phi$ which implies that the time variation of the gravitational potential $\Phi$ can be ignored.

The general relativistic equations (\ref{grper1})-(\ref{grcons2})
lead to the following equation for the matter overdensity $\delta$ \be {\ddot \delta} + 2 H {\dot \delta} + \frac{k^2}{a^2} \Phi = 0 \label{grdel1} \ee
Using the sub-Hubble approximation (\ref{drhomss}) we obtain the scale independent approximate equation (\ref{greqtim}). On the other hand, if we avoid this approximation in equation (\ref{drhom}), solve for $\Phi$ and substitute in equation (\ref{grdel1}) we obtain the following scale dependent evolution equation for $\delta$ \cite{Dent:2008ia}:
\be {\ddot \delta} + 2 H {\dot \delta} - \frac{4\pi G \rho_m \delta}{1+\xi(a,k)} =0 \label{greq-scdep} \ee
where \be \xi(a,k)=\frac{3 a^2 H(a)^2}{k^2} \label{xidef} \ee
The solution of equation (\ref{greq-scdep}) provides a much better approximation to the full linear general relativistic system (\ref{grper1})-(\ref{grper3}) up to horizon scales.  On scales larger than the horizon even equation (\ref{greq-scdep}) breaks down since on these scales, the time derivative of $\Phi$ can not be ignored.

Given the successful approximation of the solution of (\ref{greq-scdep}) to the exact linear general relativistic solution, it becomes important to construct a scale dependent parametrization that is analogous to (\ref{f0def}) and solves (approximately) (\ref{greq-scdep}) for all scales $k$. In order to construct such a parametrization we focus on the matter dominated era when most of the growth occurs and express $\xi(a,k)$ as
\be \xi(a,k)=\frac{3 H_0^2 \omm}{a k^2} \label{xidef2} \ee Equation (\ref{greq-scdep}) may be expressed in terms of the growth factor $f=\frac{d\ln \delta}{d\ln a}$ in the form
\be f' + f^2 + \(2-\frac{3}{2} \omms(a)\)f=\frac{3}{2}\frac{\omms(a)}{1+\xi(a,k)} \label{greqlna-scdep} \ee
where $'\equiv \frac{d}{d\ln a}$, and we have assumed \lcdm for $H(a)$.  Notice that the matter domination approximation is implemented explicitly only with respect to the scale dependent correction. The sub-Hubble scale independent parametrization does not explicitly set $\Omega_m(a)=1$ but its derivation implicitly assumes $\Omega_m(a)\simeq 1$ (see eg \cite{Wang:1998gt}). Thus we have used $\omms(a)$ in eq. (\ref{greqlna-scdep}) in order to make connection with the sub-Hubble scale independent parametrization $f_0(a)$.

For sub-Hubble scales $\xi(k,a)\rightarrow 0$ and equation (\ref{greqlna-scdep}) reduces to (\ref{greqlna}) whose solution is well approximated by (\ref{f0def}) with $\gamma=\frac{6}{11}$.

We now consider a small $\xi$ and look perturbatively for a solution of the form \be f(k,a)=f_0(a) (1-f_1(k,a)) \label{pertsol1} \ee Substituting (\ref{pertsol1}) in (\ref{greqlna-scdep}) and keeping only terms linear in $f_1$ and $\xi$, it is straightforward to show that \be f_1(k,a)=\xi(k,a)=\frac{3 H_{0}^{2} \omm}{a k^2} \label{f1sol} \ee where we have also used the fact that in a matter dominated universe $\delta(a)\sim a$.

This analysis may be extended to arbitrary order in perturbation theory by expanding $(1+\xi(k,a))^{-1}$ \be (1+\xi(k,a))^{-1}=\sum_{n=0}^N (-1)^n \xi(k,a)^n \ee and looking for a solution of the form \be f(k,a)=f_0(a)\sum_{n=0}^N (-1)^n f_n(k,a)^n \ee It is then straightforward to show order by order that $f_n(k,a)=\xi(k,a)$. Therefore, the parametrization \be f(k,a)=\frac{f_0(a)}{1+\xi(k,a)}=\frac{\omms(a)^\gamma}{1+\frac{3 H_0^2 \omm}{a k^2}} \label{newpar} \ee is an approximate solution of equation (\ref{greqlna-scdep}) and provides a good approximation to the solution of the general relativistic system (\ref{grper1})-(\ref{grper3}) up to horizon scales.

Alternatively, the form of $f(k,a)$ for a \lcdm background can also be deduced in a manner similar to the appendix in \cite{Wang:1998gt}. Guided by the matter-dominated case, we assume that
\be
\label{xiX}
f(k,a)=\frac{\omms(a)^\gamma}{1+\frac{X(a)}{a k^2}}
\ee
where $X(a)$ is some arbitrary time-dependent function. We next switch to $x\equiv 1-\Omega_m\(a\)$ as the time variable, assume the \lcdm evolution for $\Omega_m\(a\)$, and express the left hand side of \eqn{greqlna-scdep} as a series in $x$:
\begin{eqnarray}
\label{xseries}
\(\frac{k^2 X_0-3 H_0^2 k^2 \Omega_{m0}}{2 H_{0}^2\(1-\Omega_{m0}\)^{1/3} \Omega_{m0}^{2/3} X_0} \)x^{1/3}&&\nonumber\\
+\( \frac{k^4 X_{0}^2-9 H_0^{4}k^4\Omega_{m0}^2-15 H_{0}^4 k^2 \Omega_{m0}^2 X_1}{6 H_{0}^{4} \(1-\Omega_{m0}\)^{2/3}\Omega_{m0}^{4/3} X_{0}^2}\)x^{2/3}+O\[x\]&&\nonumber\\
\end{eqnarray}
where we have used a Taylor expansion for the function $X(x)$
\be
\label{Xpansion}
X(x)=X_0+X_1 x+O\[x^2\]
\ee

For \eqn{xiX} to be an approximate solution to \eqn{greqlna-scdep}, we can expect the leading order coefficients of the expansion \eqref{xseries} to vanish. The vanishing of the first two coefficients immediately implies that that $X_0=3 H_{0}^2\Omega_{m0}$ and $X_1=0$.  The latter fact makes it clear that any $x$ dependence of $X$ is weak (quadratic or higher). (Note that higher coefficients of \eqref{xseries} contain non-linear combinations of the coefficients of $X(x)$, making it difficult to continue this process to deduce further coefficients in \eqn{Xpansion}.) We therefore conclude that \eqn{newpar} provides a good approximation to the solution of \eqn{greqlna-scdep}.

The accuracy of our parameterization is demonstrated in Fig. 1 where we compare the form of the scale dependent parametrization \eqn{newpar} to the general relativistic numerical solution with the corresponding fit of the standard parametrization for three different scales. It is clear that up to approximatelly the Hubble scale $(k \simeq 0.001hMpc^{-1})$ is accurate at a level better than $5\%$ at least up a to redshift $z=10$.
We further quantify the accuracy of either parameterization through $\Delta$, the fractional deviation of $f$ predicted by the parameterization (i.e. \eqn{f0def} for the scale independent case and \eqn{newpar} for the scale dependent case) from the exact $f$:
\be
\label{Delta}
\Delta\equiv\frac{f_{\rm exact}-f_{\rm parametrization}}{f_{\rm exact}}
\ee

In Fig. \ref{errs} a we display $\Delta$ for the scale independent parameterization for the same scales as considered in \fig{ffits}. This can be contrasted with the corresponding $\Delta$ for the scale-dependent case shown in Fig. \ref{errs}b.

The analysis of this section has assumed a background expansion based on \lcdm. In the next section we demonstrate that the parametrization (\ref{newpar}) is also a good approximation to the general relativistic system solution in the case of a dynamical dark energy background.

\section{Scale Dependent Growth Parametrization with Dynamical Dark Energy}

It is straightforward to solve the general relativistic system (\ref{grper1})-(\ref{grper3}) in an arbitrary homogeneous dark energy background i.e., in what follows, we ignore perturbations of the dark energy component. However for a discussion of the growth of cosmological perturbations in which the dark energy component is not assumed to be homogeneous, see e.g. \cite{duttadentweiler,motashawsilk,duttamaor} and references therein. For the evolution of $w$, we consider a dark energy parametrization \cite{Chevallier:2000qy} \be w_{de}(a)=w_0 + w_1 (1-a) \label{cpl} \ee and solve the system (\ref{grper1})-(\ref{grper3}) for various values of the parameters $(w_0,w_1)$. We then fit the parametrization \be f(k,a)=\frac{\omms(a)^\gamma}{1+\alpha\frac{3 H_{0}^{2} \omm}{a k^2}} \label{newparga} \ee to the numerical solution and determine the best fit values of the parameters $(\gamma, \alpha)$ for various values of the scale $k$ and of the dark energy parameters $(w_0,w_1)$. In particular, we first fix $(w_0,w_1)$ and also the scale $k$ to a value much smaller than the horizon (eg $k\simeq 1 h Mpc^{-1}$). We then solve numerically the system (\ref{grper1})-(\ref{grper3}), find $f(k,a)$ and fit the parametrization (\ref{newparga}) to the numerical solution thus determining the best fit value of $\gamma$ (the value of $\alpha$ is irrelevant for large $k$). Next, we consider larger scales (up to $k=10^{-3} h Mpc^{-1}$) and for each value of $k$ we find the best fit value of $\alpha$ for the fixed value of $\gamma$ obtained from the sub-Hubble fit. As anticipated from the discussion of the previous section we find that the best fit value of $\alpha$ is practically independent of the scale $k$ and is equal to 1. Finally, we repeat the above steps for different values of $(w_0,w_1)$ to find the dependence of the best fit parameters $(\gamma, \alpha)$ on the dark energy properties.

We find that for values of $(w_0,w_1)$ consistent with observations \cite{Nesseris:2006er} $(\gamma, \alpha)\simeq(0.55,1)$. This is demonstrated in Fig. 3 where we show the dependence of the best fit parameter values for $(\gamma,\alpha)$ in terms of $w_0$ (for fixed $w_1=0$ in Fig. 3a) and in terms of $w_1$ (for fixed $w_0=-1.4$ in Fig. 3b). The error bars in the parameter $\alpha$ describe the $1\sigma$ variation of $\alpha$ with the scale $k$. This is clearly negligible except of the pair $(w_0,w_1)=(-1.4,1.2)$ which implies significant presence of dark energy at early times thus affecting our assumption of matter domination at early times.

\begin{figure}
	\epsfig{file=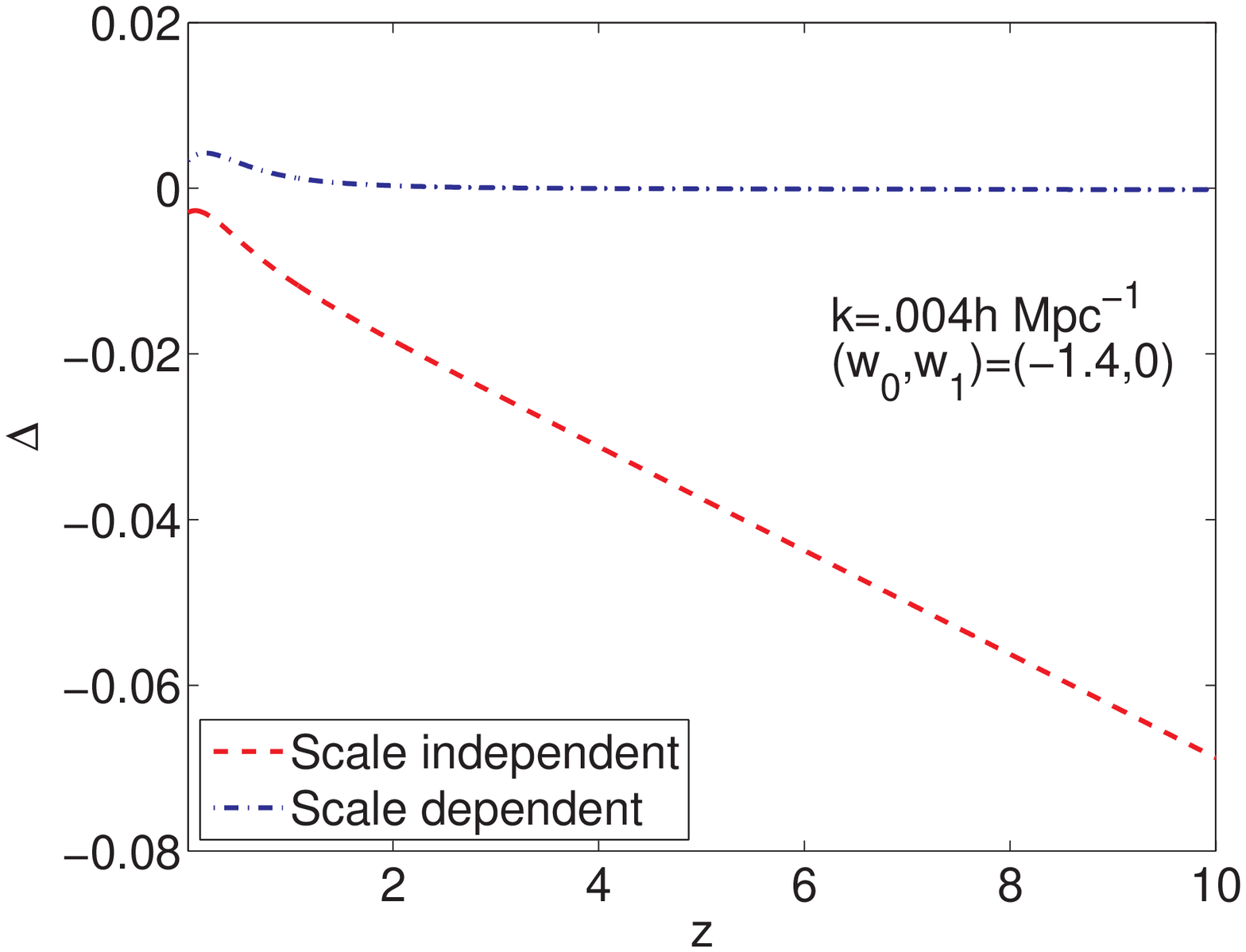,height=55mm}
	\epsfig{file=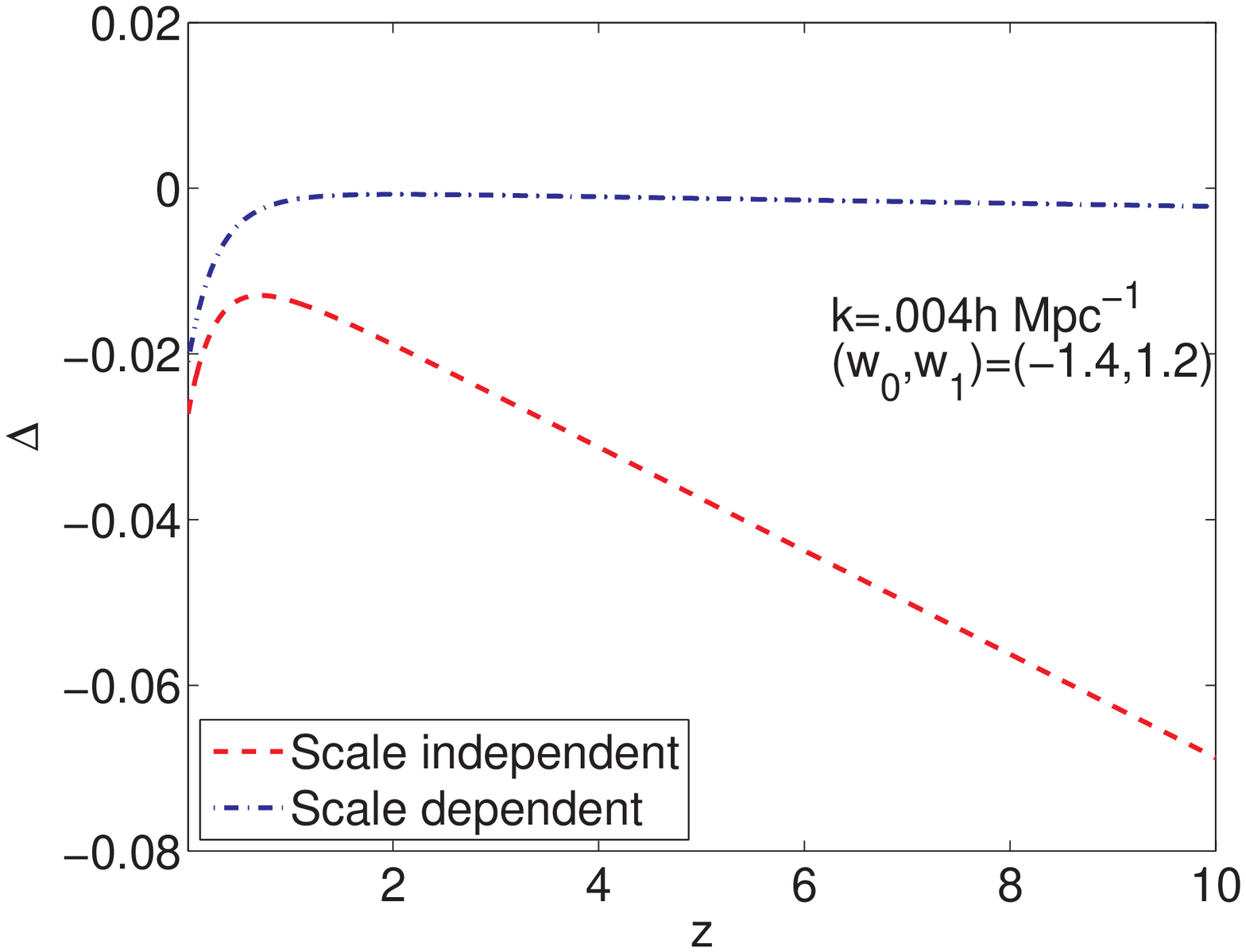,height=55mm}
	\caption
	{\label{w0w1}
$\Delta$ for the scale independent  and scale dependent  parameterizations at a scale of k=.004h Mpc$^{-1}$ for the ($w_0,w_1$) choices shown. }
\end{figure}

In \fig{w0w1} we plot $\Delta$ for the scale dependent and independent parameterizations for the values $(w_0,w_1)=(-1.4,0)$ (top frame) and $(w_0,w_1)=(-1.4,1.2)$ (bottom frame).

\section{Conclusion - Discussion}

We have found a scale dependent parametrization of the growth function $f=\frac{d\ln \delta}{d \ln a}$ that is free from the sub-Hubble approximation of the standard parametrization (\ref{f0def}). Our parametrization described by equation (\ref{newpar}) approximates very well the full linear general relativistic solution up to horizon scales and differs significantly from the standard parametrization on scales larger than about $50 h^{-1}Mpc$.

An important extension of our work is the derivation of the best fit parameter values $(\gamma,\alpha)$ in the context of modified gravity models. It is also interesting to investigate whether a more general scale dependent parametrization is required to describe the growth function on large scales in modified gravity theories.

The mathematica files used for the production of the figures may be downloaded from http://leandros.physics.uoi.gr/newpar.zip

\section*{Acknowledgements}
This work was supported by the European Research and
Training Network MRTPN-CT-2006 035863-1 (UniverseNet). JBD and SD were supported in part by U.S. DoE grant DE-FG05-85ER40226.

\end{document}